\documentclass[10pt,twocolumn]{article}
\usepackage[letterpaper,margin=0.75in,columnsep=0.3in]{geometry}
\usepackage[T1]{fontenc}
\usepackage{libertine}
\usepackage[libertine]{newtxmath}
\usepackage{microtype}
\usepackage{graphicx}
\usepackage{booktabs}
\usepackage{tabularx}
\usepackage{array}
\usepackage{enumitem}
\usepackage[font=small,labelfont=bf]{caption}
\usepackage{xcolor}
\usepackage{xurl}
\usepackage[colorlinks=true,linkcolor=black,citecolor=blue!50!black,urlcolor=blue!50!black]{hyperref}
\setlist{nosep,leftmargin=*}
\usepackage{titlesec}
\titleformat*{\section}{\large\sffamily\bfseries}
\titleformat*{\subsection}{\normalsize\sffamily\bfseries}
\newcommand{\COtwo}{CO\textsubscript{2}}
\graphicspath{{img/}}
\title{\textbf{EnRoute: Iterative Design and Evaluation of a Mobile App for Lower-Carbon Fashion and Transportation Choices}}
\author{Angela Busheska\\Lafayette College, Easton, Pennsylvania, USA}
\date{}

\begin{document}
\maketitle

\begin{abstract}
EnRoute is an iOS application designed to support more sustainable everyday choices in transportation and fast fashion consumption. We describe three design iterations and the user research and field evaluations that shaped each one. Interviews with 15 people from 12 countries informed Version I, which combined errand route planning with basic fashion guidance. A survey of 1,200 people in five countries found much stronger interest in sustainable shopping (78\%) than in switching to lower-carbon transport (46\%); respondents most often named limited public transport as the main barrier to lower-carbon travel. We therefore narrowed the focus to fashion. Version II added item-level garment scanning with a manual-entry fallback, a Sustainability Coefficient that rates items, and recommendations of more sustainable alternatives; in a two-week evaluation with 3,021 participants in eight countries, participants randomly assigned to receive a short tutorial and educational prompts viewed more recommended alternatives and reported larger gains in purchase intent than those without it. Version III shifted from ``buy better'' to ``buy less,'' adding a wardrobe inventory, outfit matching, and a ``Real Price'' that combines cost per wear with estimated environmental and social costs. In a four-week pre/post pilot with 500 participants, self-reported monthly garment purchases fell from 2.9 to 1.9. We report descriptive results, discuss the substantial limitations of these uncontrolled and self-reported measures, and outline what the iterations suggest about designing for sustainable consumption.
\end{abstract}


\section{Introduction}
The capital of North Macedonia, Skopje, has been listed by the United Nations Environment Programme among the European cities with the highest concentrations of fine particulate matter (PM\textsubscript{2.5}) \cite{unep}. According to an analysis by the Ministry of Environment and Physical Planning, traffic accounts for up to a quarter of the city's emissions and is one of the major causes of its air pollution \cite{dodovski}. Transport emissions are not only a local problem: in 2016 they accounted for roughly a quarter of global \COtwo{} emissions from fuel combustion and were growing faster than those of any other sector \cite{wri}. Under current policies, global transport emissions are projected to rise by about 60\% between 2015 and 2050 \cite{planete}.

Fast fashion is a second, largely hidden source of emissions. The fashion industry is estimated to produce 8 to 10\% of global \COtwo{} emissions \cite{niinimaki}. Brands such as Shein, Zara, and H\&M manufacture in countries with low labor costs, where much of the environmental damage falls on local communities rather than on the companies or their customers \cite{overconsumption}. In North Macedonia, as in many countries, regulation of this industry is limited.

Accessible, transparent information about one's own contribution to pollution is often hard to find, which can make it difficult to act. At the same time, people who do try to live more sustainably are vulnerable to greenwashing, where companies make misleading environmental claims and consumers end up supporting them in good faith \cite{tsui}.

To help people make lower-carbon choices in transportation and fashion, we developed EnRoute, an iOS application. This paper describes its three design iterations, the user research and evaluations behind each, and what we learned. All results reported here are descriptive and come from studies run during product development; we discuss their limitations in Section~\ref{sec:limitations}.

\section{Related Work}
Sustainable HCI has a long history of designing technology for environmentally responsible behavior \cite{disalvo}. Much of this work draws on persuasive technology \cite{fogg} and on theory-driven strategies for supporting behavior change in everyday life \cite{consolvo}, as well as on eco-feedback, which makes otherwise invisible environmental impacts visible to the people who cause them \cite{froehlich2010}. In transportation, UbiGreen showed that a mobile tool could track green commuting habits and give ambient feedback on them \cite{ubigreen}. At the same time, critics have argued that focusing narrowly on persuading individuals can overlook the structural conditions that shape what people are able to do \cite{brynjarsdottir}. Our Version I survey, in which respondents most often named inadequate public transport as the main barrier to low-carbon travel, echoes this concern. EnRoute extends this line of work to fashion consumption, where the environmental cost of a purchase is largely invisible at the point of sale \cite{niinimaki}.

\section{Version I: User Research and Design}
\subsection{Interviews}
\begin{figure}[t]
\centering
\includegraphics[width=\columnwidth]{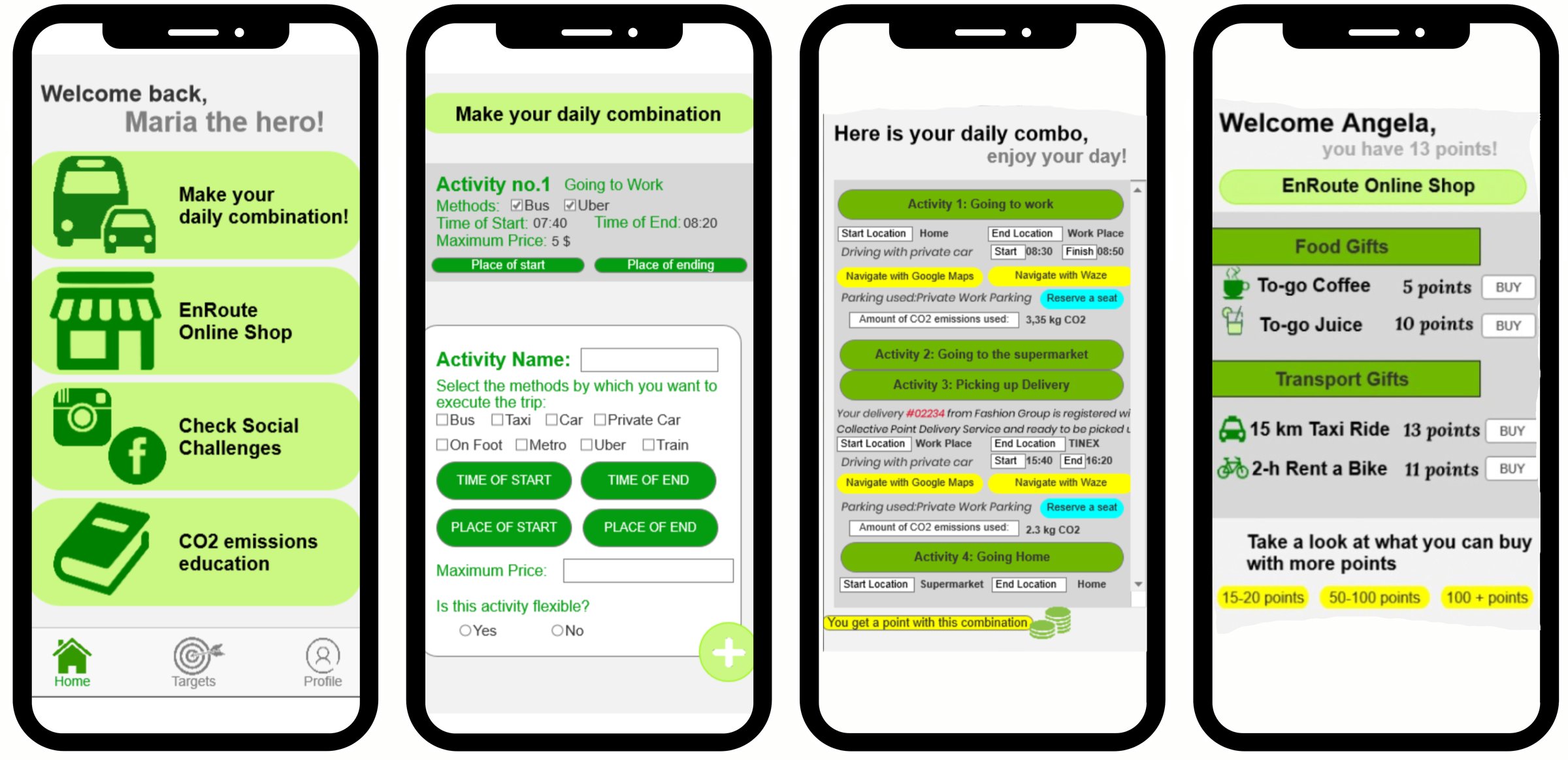}
\caption{Initial EnRoute prototypes used in the interviews, including public transport scheduling and sustainable shopping screens.}
\label{fig:v1proto}
\end{figure}

Our goal was to understand how prospective users from different socio-cultural backgrounds would perceive and use an app aimed at reducing everyday carbon emissions. Fifteen people (P1 to P15) from twelve countries took part, recruited through personal contacts, online networks, and referrals. Interviews were structured around early mock-ups (Figure~\ref{fig:v1proto}).

\begin{table}[t]
\centering\small
\caption{Feature ideas raised by interview participants (N = 15), with the number of participants mentioning each and example quotes.}
\label{tab:interviews}
\begin{tabularx}{\columnwidth}{>{\raggedright\arraybackslash}p{0.27\columnwidth} c >{\raggedright\arraybackslash}X}
\toprule
Feature / idea & $n$ & Example quotes \\
\midrule
Public transport integration & 10 & ``Include bus schedules and train times to simplify logging.'' (P1); ``Integrate scooters, bikes, and public transit.'' (P2) \\
Social / community challenges & 11 & ``Seeing friends' progress makes it more fun.'' (P3); ``Compete with others for a good cause.'' (P1) \\
Gamified rewards (points, leaderboard) & 12 & ``Collecting points is brilliant.'' (P2); ``Leaderboards make it exciting to compete with friends.'' (P4) \\
Fast fashion awareness & 7 & ``Help users know which brands are harming or helping the environment.'' (P5); ``Include sustainability ratings for brands.'' (P6) \\
Streaks or personal bests & 6 & ``Streaks are motivating but can also cause guilt.'' (P5); ``Show users their personal best for a day or week.'' (P2) \\
Partner rewards & 9 & ``Partner with Starbucks for redeemable points.'' (P2); ``Offer discounts at eco-friendly stores.'' (P6) \\
Education on \COtwo{} emissions & 8 & ``Include tips to be a better advocate.'' (P5); ``Education is critical, especially in Guatemala.'' (P7) \\
Daily activity logging & 7 & ``Add one-off trips for better flexibility.'' (P8); ``Simplify how activities are logged.'' (P1) \\
\bottomrule
\end{tabularx}
\end{table}

\subsection{Key Insights}
We grouped participants' suggestions into eight feature categories (Table~\ref{tab:interviews}). Beyond engagement mechanics such as rewards and challenges, participants' suggestions clustered around two domains: transportation and sustainable fashion. Participants saw transport as a major part of their daily footprint and wanted practical, low-carbon alternatives: real-time public transport schedules, carpooling, and safe biking routes. Discussions of fashion centered on education and transparency; participants wanted to know how sustainable the brands they buy from really are, and several suggested brand ratings and buying guides.

\subsection{Feature Design}
Based on these insights we built two core features.

\textbf{Sustainable travel.} The ``Make Your Daily Combination'' feature combines several errands (for example, commuting and grocery shopping) into a single route. Using the user's preferences (budget, acceptable walking distance) and the current weather, it estimates the \COtwo{} footprint of the user's usual routine and compares it with that of an optimized itinerary.

\textbf{Sustainable shopping.} Users can scan a product label to see sustainability information about the brand, including its carbon footprint, labor practices, and water use. Because many people assume sustainable clothing is expensive, the feature also suggests budget-friendly alternatives that match the user's preferences.

\begin{figure}[t]
\centering
\includegraphics[width=\columnwidth]{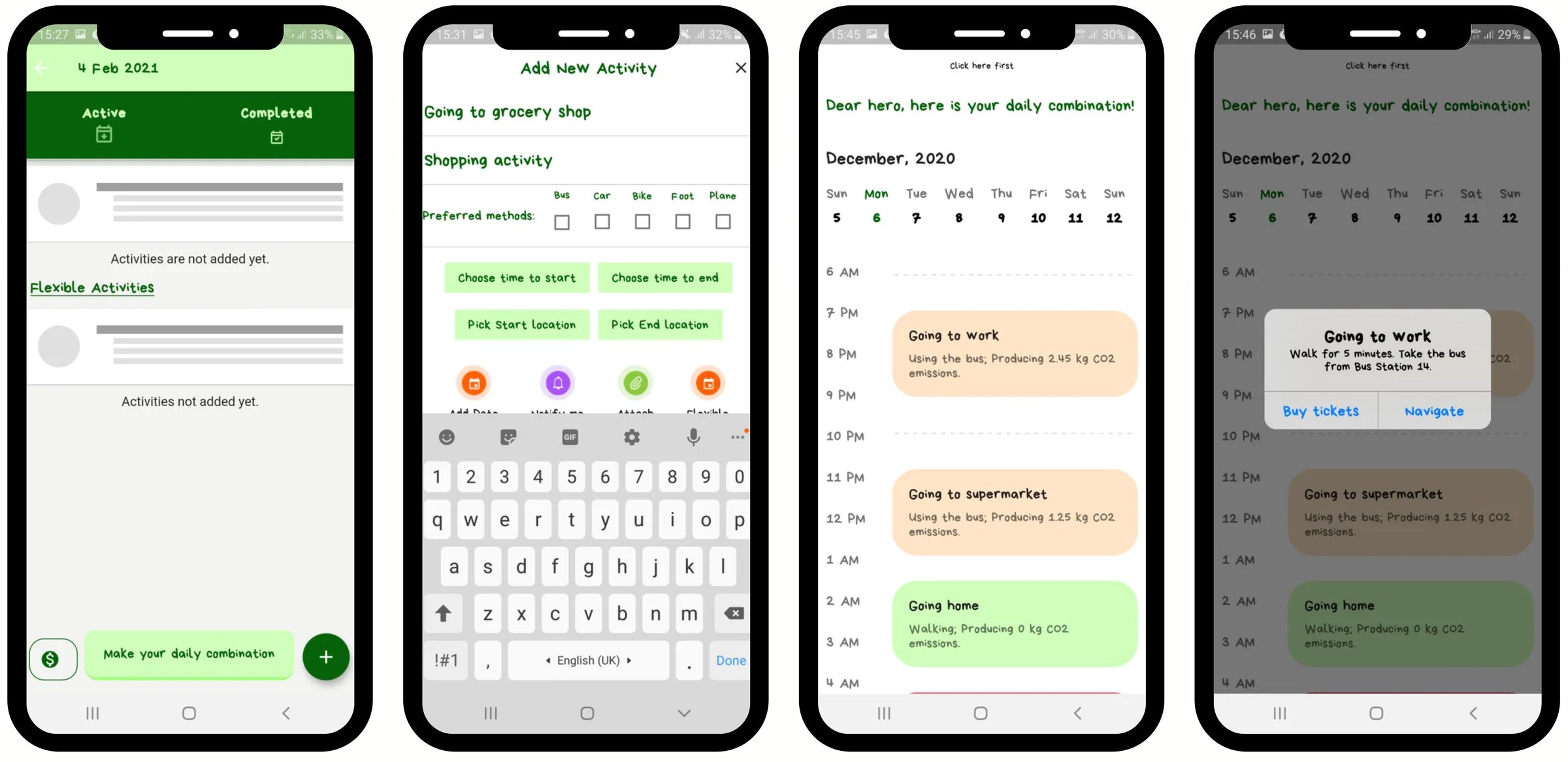}
\caption{Redesigned ``Make Your Daily Combination'' feature, with a simplified pop-up for selecting activities and an optimized route to reduce \COtwo{} emissions.}
\label{fig:v1redesign}
\end{figure}

\begin{figure}[t]
\centering
\includegraphics[width=\columnwidth]{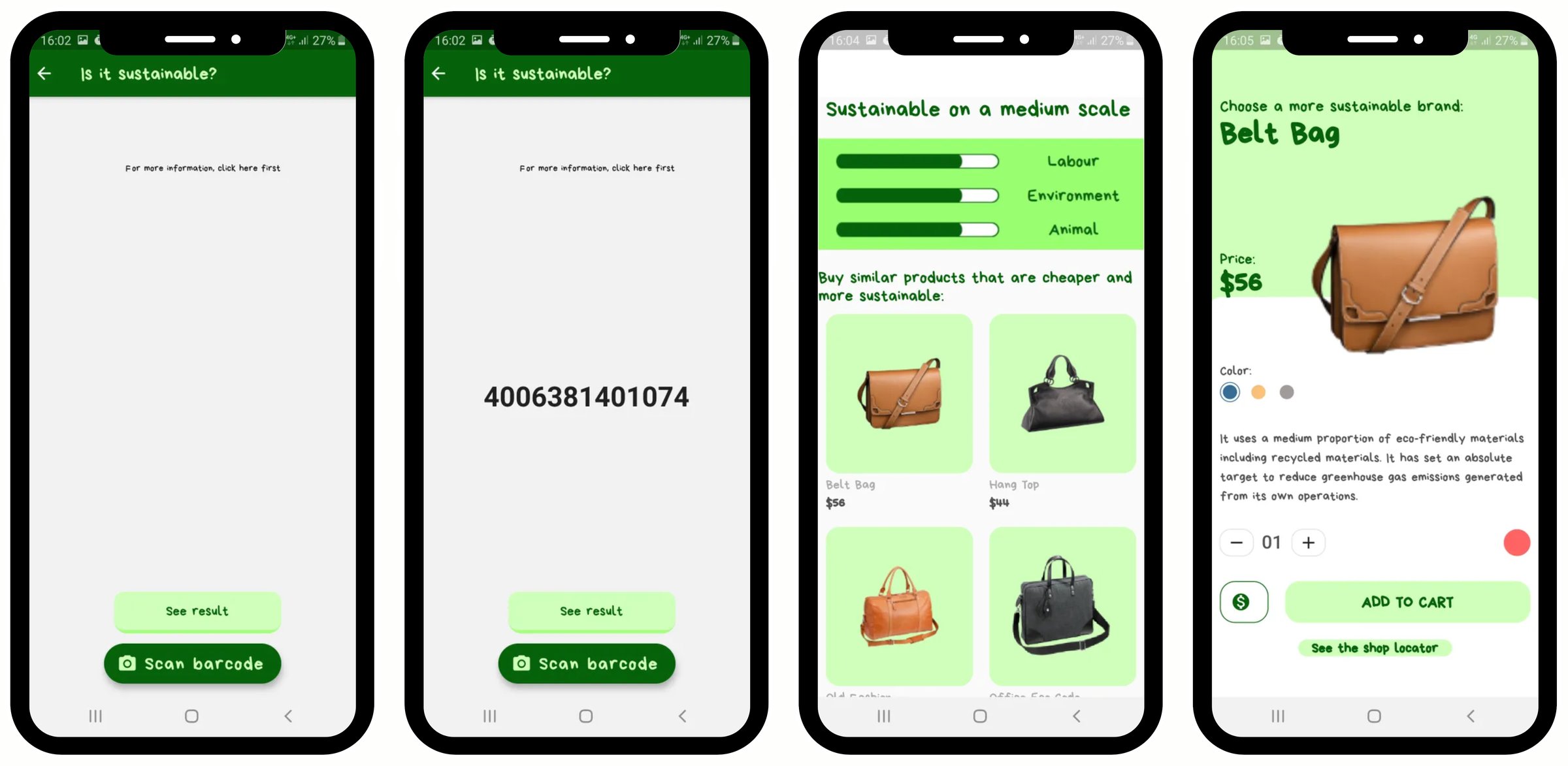}
\caption{Redesigned fashion interface for scanning clothing items and discovering more sustainable alternatives.}
\label{fig:v1fashion}
\end{figure}

We also collected feedback on usability. Participants found the calendar view in ``Make Your Daily Combination'' cluttered and unintuitive, so we moved activity selection into a separate pop-up with a simpler layout (Figure~\ref{fig:v1redesign}). Figure~\ref{fig:v1fashion} shows the fashion tab.

\section{Version I: Survey}
We surveyed 1,200 people in five countries about their willingness to adopt lower-carbon travel (for example, public transport or carpooling) and their interest in shopping sustainably, particularly for clothing. Interest was rated on a 1 to 5 scale. Results are summarized in Table~\ref{tab:v1}.

\begin{table}[t]
\centering\small
\caption{Version I survey results (N = 1,200). Percentages are the share of respondents reporting high interest.}
\label{tab:v1}
\begin{tabular}{lcc}
\toprule
 & Sustainable & Lower-carbon \\
 & shopping & transport \\
\midrule
High interest & 78\% & 46\% \\
Mean interest (1 to 5) & 4.2 (SD 0.7) & 3.3 (SD 1.0) \\
Main barrier & Perceived cost & Limited transit \\
\bottomrule
\end{tabular}
\end{table}

Interest in sustainable shopping was considerably higher than interest in switching to lower-carbon transport (78\% vs.\ 46\% reporting high interest; mean 4.2 vs.\ 3.3). In open-ended responses, respondents named inadequate public transport and affordability as the main barriers to low-carbon travel, and high perceived cost and lack of transparency about brands as the main barriers to sustainable shopping.

Because infrastructure is largely outside an app's control, we decided to focus the next iteration on fashion, where respondents were more motivated and where better information and cost comparisons could plausibly make a difference.

\section{Version II: Design}
\textbf{Preference input.} Users enter shopping preferences, including budget, location, and openness to new brands. The app uses these to suggest alternative items with higher Sustainability Coefficients.

\textbf{Scanning with manual fallback.} Users can scan clothing items in the app. When a scan recognizes the item, its brand and product details appear immediately. When it does not, the app asks the user to enter the brand and material composition by hand, so an assessment is still possible when visual recognition fails.

\begin{figure}[t]
\centering
\includegraphics[width=0.45\columnwidth]{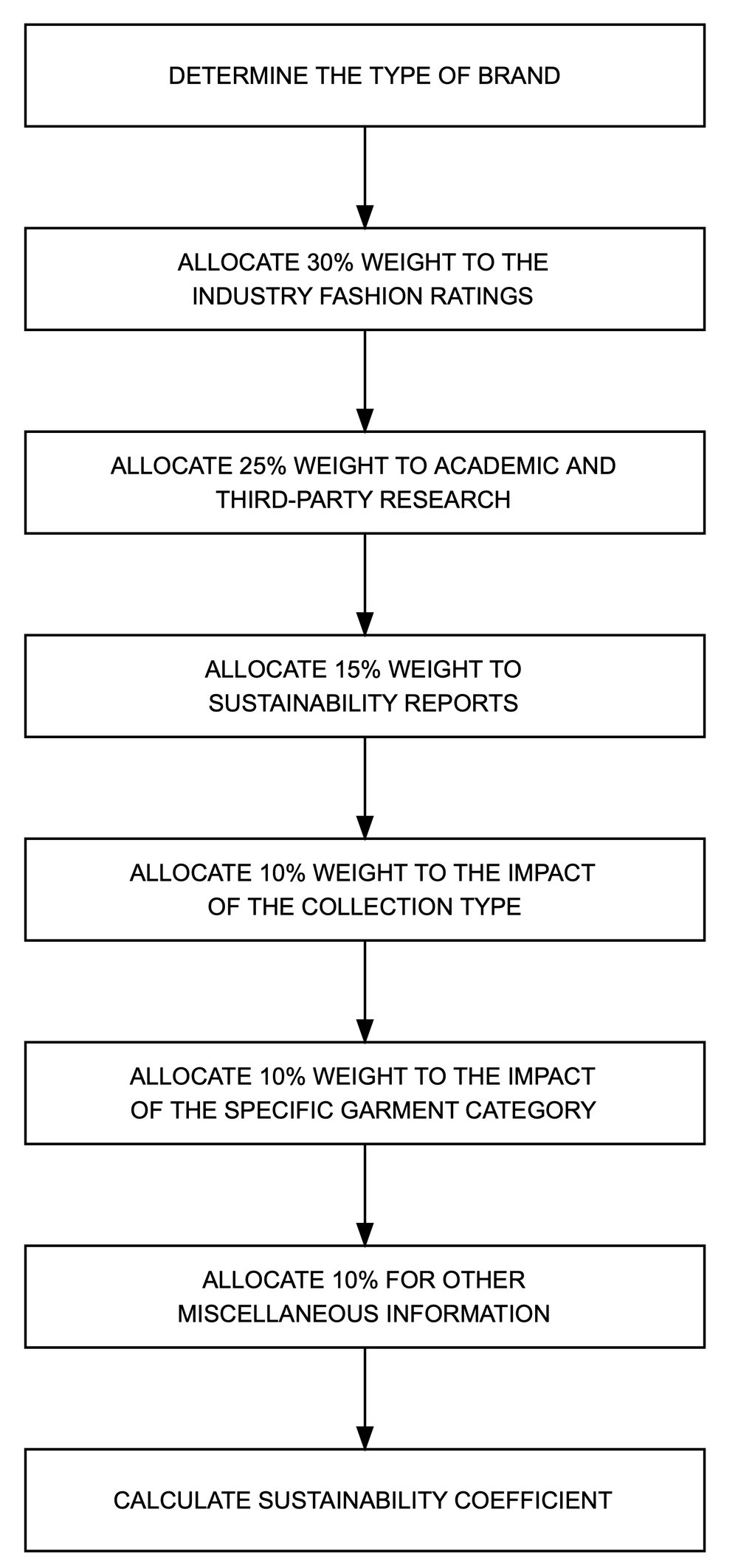}
\caption{How the Sustainability Coefficient is calculated, from determining the brand type through the weighted combination of six information sources.}
\label{fig:coef}
\end{figure}

\textbf{Sustainability Coefficient.} Each item receives a coefficient, where higher values indicate a more sustainable item. It draws on information about material composition, carbon footprint, and production practices, weighted by source (Figure~\ref{fig:coef}):
\begin{itemize}
\item 30\% industry sustainability ratings
\item 25\% academic and third-party research
\item 15\% company sustainability reports
\item 10\% impact of the collection type (for example, a seasonal fast fashion line versus a core collection)
\item 10\% impact of the garment type
\item 10\% other information
\end{itemize}
These weights reflect design judgment rather than empirical calibration; validating them is future work.

\begin{figure}[t]
\centering
\includegraphics[width=\columnwidth]{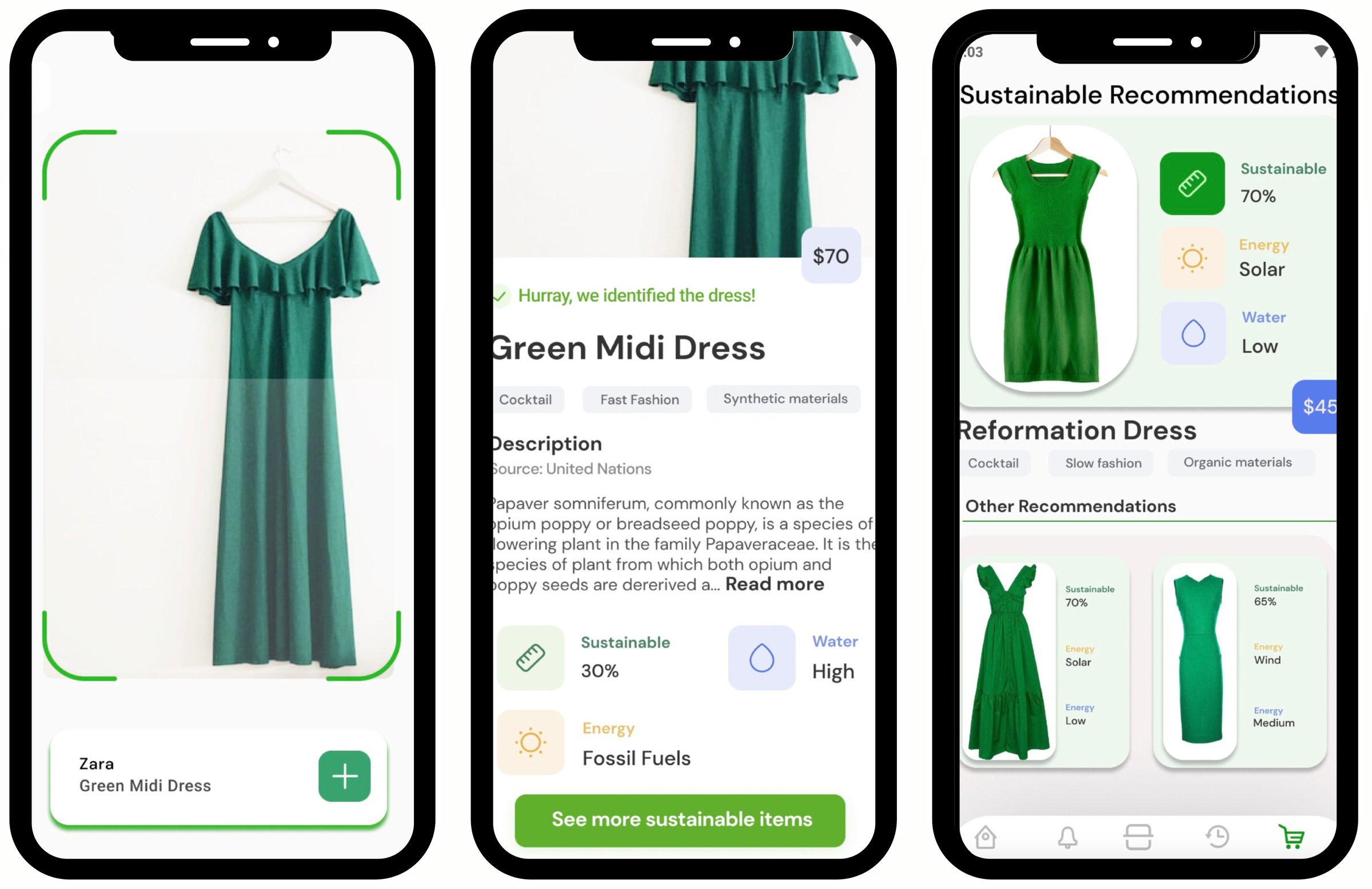}
\caption{Version II recommendation interface, showing more sustainable alternatives ranked by Sustainability Coefficient and user preferences.}
\label{fig:v2ui}
\end{figure}

\textbf{Recommendations.} A ranking step filters alternatives by sustainability score, budget, and geographic availability and presents the top options (Figure~\ref{fig:v2ui}).

\section{Version II: Evaluation}
\subsection{Study Design}
We evaluated Version II with 3,021 participants in eight countries over two weeks. The study examined (1) the dual input mechanism (scanning and manual entry), (2) how clear and trustworthy participants found the Sustainability Coefficient, and (3) whether the recommendations changed participants' intent to buy sustainable clothing.

Participants were recruited through social media ads, sustainability forums, and local community partnerships; the only inclusion criterion was a smartphone with a camera. Before using the app, participants completed a short questionnaire on their attitudes toward sustainable shopping and how often they bought eco-friendly products. After two weeks they completed a post-study survey on trust in the scoring, usability, and purchase intent.

To examine the effect of in-app education, participants were randomly assigned at sign-up to one of two groups of roughly equal size:
\begin{itemize}
\item \textbf{Group A (no tutorial):} saw the coefficient and recommendations, with only a single onboarding screen and no further explanation.
\item \textbf{Group B (tutorial and prompts):} viewed a two-minute interactive tutorial on how the coefficient works at first launch, and saw short tips (for example, ``Why does cotton have a lower environmental impact than polyester?'') whenever they scanned a new item.
\end{itemize}

\subsection{Measures}
\begin{itemize}
\item \textbf{Scan recognition accuracy:} share of scans in which the app correctly identified the brand and garment type.
\item \textbf{Manual input usage:} share of participants who entered item details by hand at least once when scanning failed.
\item \textbf{Clarity of and trust in the coefficient:} post-study ratings of how clear and trustworthy participants found it.
\item \textbf{Alternatives viewed:} number of recommended items a participant opened, saved, or clicked.
\item \textbf{Change in purchase intent:} pre- and post-study ratings (1 to 5) of how likely participants were to buy sustainable clothing in the near future.
\item \textbf{Open-ended responses:} post-study descriptions of obstacles to and motivations for sustainable shopping.
\end{itemize}

\subsection{Results}
Results by group are shown in Table~\ref{tab:v2}. Scan recognition accuracy was 84.2\% overall and nearly identical across groups, as expected, since the tutorial did not change the scanner. Overall, 69.8\% of participants used manual entry at least once, suggesting that the fallback was important for coverage.

Participants who received the tutorial and prompts (Group B) reported higher clarity of the coefficient (79\% vs.\ 72\% rating it 4 or 5) and somewhat higher trust (71\% vs.\ 65\%). They also viewed more recommended alternatives (3.1 vs.\ 1.8 items on average) and reported a larger increase in purchase intent (+1.3 vs.\ +0.9 points). Purchase intent rose in both groups, but because there was no group without the app, we cannot attribute that overall increase to the app itself.

\begin{table}[t]
\centering\small
\caption{Version II results from the two-week evaluation (N = 3,021), by group.}
\label{tab:v2}
\begin{tabular}{lcc}
\toprule
Measure & Group A & Group B \\
 & (no tutorial) & (tutorial) \\
\midrule
Scan recognition accuracy & 83.7\% & 84.6\% \\
Used manual input at least once & 68.5\% & 71.1\% \\
Rated coefficient clear (4 or 5) & 72\% & 79\% \\
Moderate to high trust in coefficient & 65\% & 71\% \\
Alternatives viewed (mean per user) & 1.8 & 3.1 \\
Change in purchase intent (1 to 5) & +0.9 & +1.3 \\
\bottomrule
\end{tabular}
\end{table}

\section{Version III: Design}
\subsection{Motivation}
Versions I and II encouraged people to buy more sustainable items. But not buying at all is often the most effective way to reduce the footprint of clothing \cite{sandin}, and extending the active life of garments by nine months has been estimated to reduce their carbon, water, and waste footprints by 20 to 30\% \cite{wrap}. Version III therefore focused on helping people reuse and restyle what they already own before buying something new, in line with the slow fashion movement.

\subsection{Features}
\begin{figure}[t]
\centering
\includegraphics[width=\columnwidth]{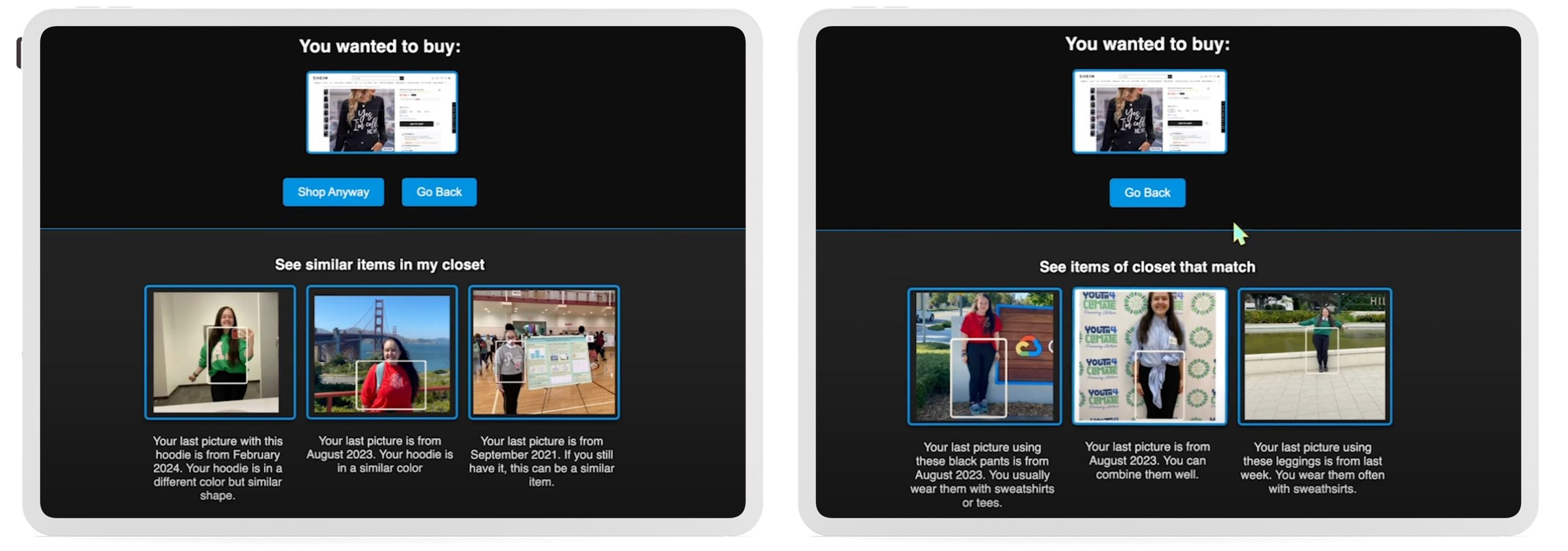}
\caption{(a) Notification that the user already owns similar items; (b) outfit matching, showing how a potential purchase pairs with items already in the user's wardrobe.}
\label{fig:v3closet}
\end{figure}

\textbf{Wardrobe inventory.} Users register items they already own by scanning or manual entry. When they consider buying something new, the app first shows similar items already in their wardrobe (Figure~\ref{fig:v3closet}a).

\begin{figure}[t]
\centering
\includegraphics[width=\columnwidth]{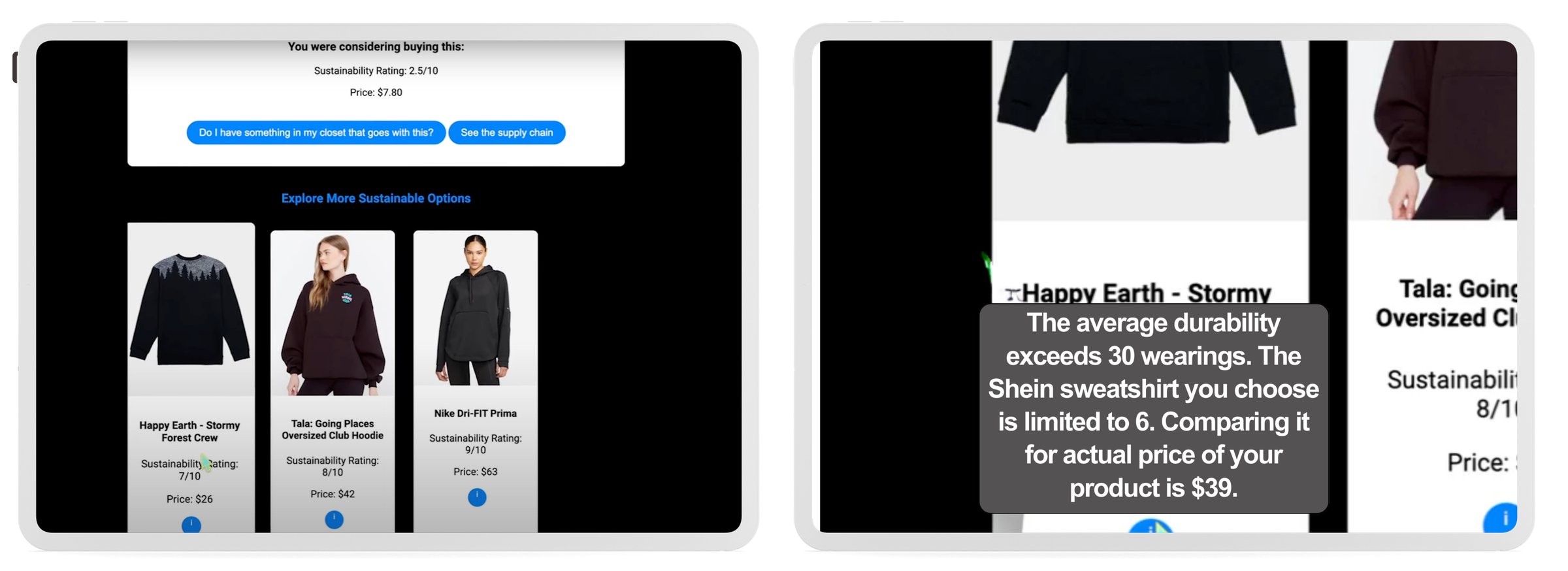}
\caption{(a) Suggestions for more sustainable clothing choices; (b) Real Price of an item compared with more sustainable alternatives, based on durability, environmental impact, and cost.}
\label{fig:v3price}
\end{figure}

\textbf{Real Price.} Drawing on life-cycle assessment principles \cite{iso14040,lin}, Version III estimates a Real Price for each garment that combines:
\begin{itemize}
\item \textbf{Cost per wear:} price divided by the expected number of wears before the garment wears out, based on typical fabric durability and quality.
\item \textbf{Environmental cost:} estimated carbon and water footprint, spread across the garment's lifespan.
\item \textbf{Social impact:} a factor based on the brand's transparency and labor conditions.
\end{itemize}
A higher Real Price means lower durability or greater environmental and social cost per wear (Figure~\ref{fig:v3price}b). Item pages also link to basic supply chain information about the brand (Figure~\ref{fig:v3price}a).

\textbf{Outfit matching.} Going beyond finding similar items, outfit matching checks how a potential purchase would combine with what the user owns, using color, style, fabric, and season tags (for example, finding tops that go with a pair of trousers). Seeing how a new item fits with what they own helps users avoid ``standalone'' purchases (Figure~\ref{fig:v3closet}b).

\begin{figure}[t]
\centering
\includegraphics[width=\columnwidth]{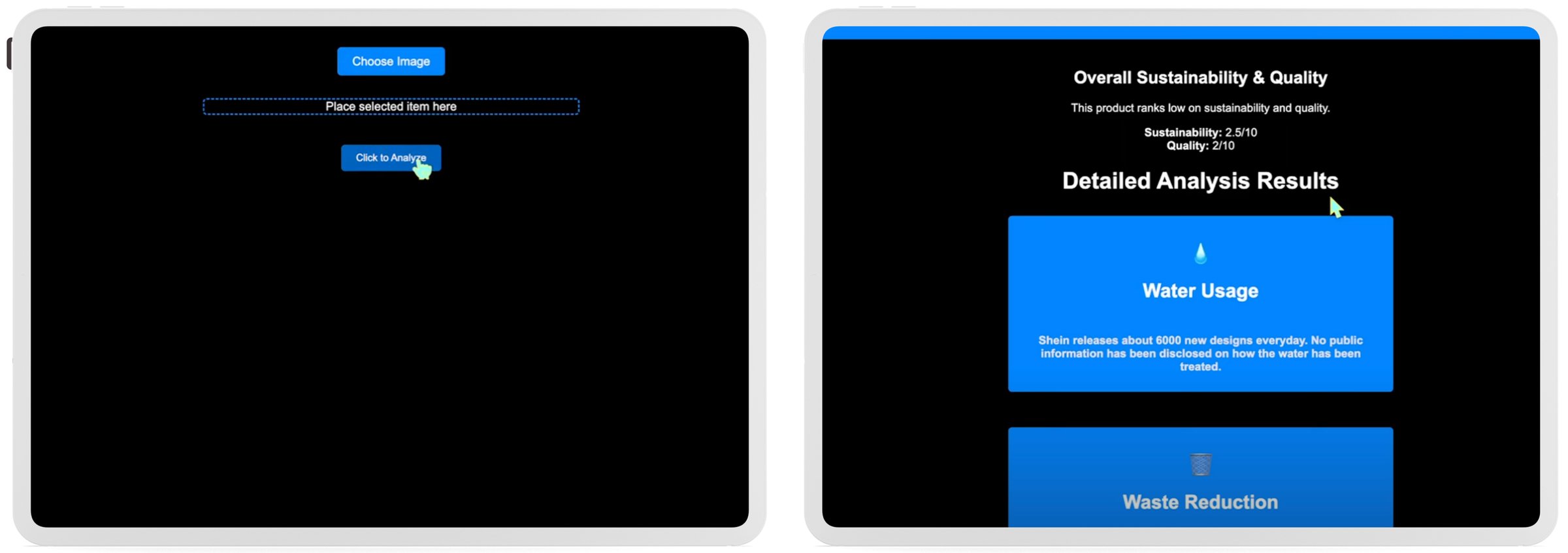}
\caption{Item analysis: users upload a photo of a garment and receive an analysis of water use, waste, and overall sustainability and quality.}
\label{fig:v3analysis}
\end{figure}

\textbf{Item analysis.} Users can also upload a photo of a garment to get a detailed analysis including water use, waste, and overall sustainability and quality ratings (Figure~\ref{fig:v3analysis}).

\section{Version III: Evaluation}
\subsection{Study Design}
We ran a four-week pre/post pilot study to explore whether the ``buy less'' features were associated with fewer new garment purchases, and how participants rated Version III's usability and trustworthiness. Participants were required to own a compatible smartphone and to be willing to use the wardrobe scanning and outfit matching features. The final sample of 500 participants was 65\% female, 33\% male, and 2\% non-binary, with a mean age of 21.4 years. There was no control group.

\subsection{Measures}
We combined self-reported survey measures with logged app usage:
\begin{itemize}
\item \textbf{Monthly purchases:} self-reported number of new garments bought in the month before the study and during the four-week study period.
\item \textbf{Outfit matching uses:} number of times a participant checked a potential purchase against their wardrobe.
\item \textbf{Real Price views:} number of times a participant viewed or updated an item's Real Price.
\item \textbf{Supply chain access:} share of participants who viewed supply chain information at least once.
\item \textbf{Sustainability awareness:} self-rated awareness of sustainability (1 to 5).
\item \textbf{Trust in app data:} combined rating (1 to 5) of confidence in the accuracy and usefulness of EnRoute's data.
\item \textbf{Open-ended responses:} satisfaction, perceived credibility, and barriers to using the features.
\end{itemize}

\subsection{Results}
Results are summarized in Table~\ref{tab:v3}. Self-reported monthly garment purchases fell from 2.9 before the study to 1.9 during it, a decrease of 1.0 item (34\%). Several participants said outfit matching and Real Price made them ``rethink each purchase.''

Some participants, particularly those with large wardrobes, initially overlooked outfit matching, but it gained traction once they saw how it could prevent ``standalone pieces.'' By the end of the study the median participant had used it six times. Several described seeing possible outfit combinations as feeling ``like shopping my own closet.''

Over 75\% of participants viewed the Real Price of at least one potential purchase, and 42\% viewed it more than five times.

\begin{table}[t]
\centering\small
\caption{Version III results from the four-week pilot (N = 500). Trust, outfit matching, Real Price views, and supply chain access relate to the app itself, so they have no pre-study baseline.}
\label{tab:v3}
\begin{tabular}{lccc}
\toprule
Measure & Before & During & Change \\
\midrule
Monthly purchases (mean) & 2.9 & 1.9 & $-1.0$ \\
Sustainability awareness (1 to 5) & 3.2 & 3.9 & +0.7 \\
Trust in app data (1 to 5) & n/a & 4.0 & n/a \\
\midrule
Outfit matching uses (median) & n/a & 6 & n/a \\
Real Price views (mean) & n/a & 5.1 & n/a \\
Viewed supply chain info & n/a & 58\% & n/a \\
\bottomrule
\end{tabular}
\end{table}

\section{Ongoing Work: Supply Chain Transparency}
\begin{figure}[t]
\centering
\includegraphics[width=\columnwidth]{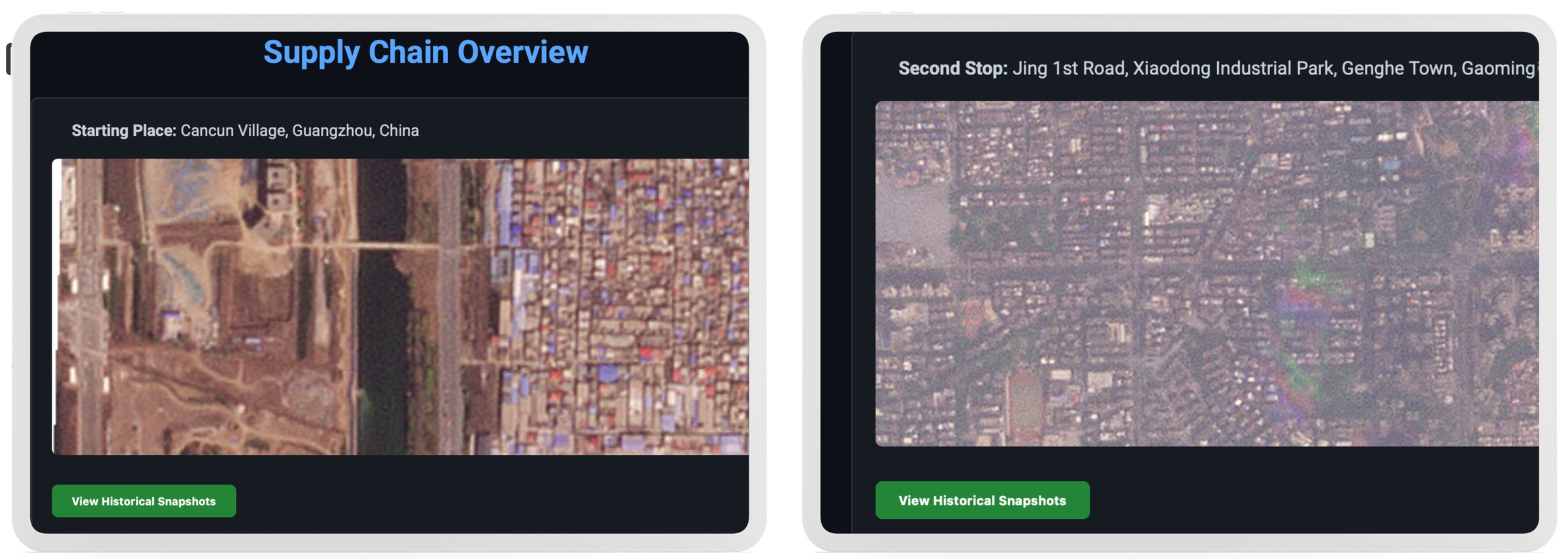}
\caption{Prototype supply chain visualization mapping stages of a garment's production onto satellite imagery.}
\label{fig:supply}
\end{figure}

France's legislative efforts to curb fast fashion \cite{dw} raise a practical question: how can one tell whether a garment was produced under fast fashion practices? Building on the basic supply chain links in Version III, we are prototyping a visualization that maps the stages of a garment's supply chain (raw material sourcing, factories, distribution centers, and retail) onto satellite imagery (Figure~\ref{fig:supply}). The aim is to give consumers and regulators a data-driven basis for distinguishing ``fast'' from ``slow'' fashion. This map-based visualization is at an early stage and has not yet been evaluated.

\section{Discussion}
Across three iterations (Table~\ref{tab:summary}), EnRoute's design moved from broad carbon tracking toward a narrower focus on clothing, where participants reported more motivation. Version I showed strong interest in sustainable shopping but much weaker willingness to change how people travel, with respondents most often naming limited public transport as the barrier. In Version II, a short tutorial and in-context explanations were associated with greater clarity, trust, and engagement with the Sustainability Coefficient, which suggests that explaining a metric may increase how clear and useful people find it. In Version III, reuse-focused features were associated with a self-reported reduction in new purchases.

\begin{table}[t]
\centering\small
\caption{Summary of the three versions.}
\label{tab:summary}
\begin{tabularx}{\columnwidth}{l >{\raggedright\arraybackslash}X >{\raggedright\arraybackslash}X >{\raggedright\arraybackslash}X}
\toprule
 & Focus & Key features & Main findings \\
\midrule
I & Transport and fashion & Route combination; brand scanning & High interest in shopping; transport limited by infrastructure \\
II & Sustainable shopping & Sustainability Coefficient; recommendations & Tutorial linked to higher clarity, trust, and engagement \\
III & Buying less & Outfit matching; Real Price & Self-reported purchases fell 34\% \\
\bottomrule
\end{tabularx}
\end{table}

\subsection{Limitations}
\label{sec:limitations}
These findings should be read as early, exploratory evidence. The evaluations were run during product development rather than as controlled experiments, and they share several limitations.

\textbf{Study design.} Version I was a survey of intentions, not of behavior. Version II randomized participants to tutorial conditions, but had no condition without the app, so increases common to both groups cannot be attributed to EnRoute. Version III had no control group at all, so the reduction in purchases may partly reflect seasonal shopping patterns, novelty, or participants' awareness of being studied.

\textbf{Measurement.} Purchase counts, intent, attitudes, and trust were self-reported and may be inflated by social desirability. We report descriptive statistics only; future work should report full inferential analyses with effect sizes and confidence intervals.

\textbf{Samples.} Participants were recruited through social media, sustainability forums, personal networks, and community partnerships, and are likely to be more interested in sustainability than the general population. The Version III sample was young (mean age 21.4).

\textbf{Metrics.} The weights in the Sustainability Coefficient and the components of the Real Price are design choices that have not been empirically validated, and both depend on corporate disclosures that vary in completeness and reliability.

\textbf{Structural limits.} As critics of persuasive sustainability technology have argued \cite{brynjarsdottir}, large shifts in transportation habits are likely to require systemic change, such as better public infrastructure, beyond what an app can provide.

Future work should include controlled studies with objective purchase data, longer follow-up to test whether effects persist, validation of the coefficient against independent life-cycle data, and collaboration with local governments and manufacturers to obtain verifiable supply chain information.

\section{Conclusion}
Through three development cycles, EnRoute moved from tracking emissions toward helping people buy better and, ultimately, buy less. Explained, transparent metrics and features grounded in what people already own were the most promising directions we found. While our evaluations are preliminary, they suggest that granular, individualized information about the true cost of clothing is a promising direction for technology that supports more sustainable consumption.

\section*{Acknowledgments}
This project was supported by a National Geographic Young Explorers Grant and the Shawn Mendes Foundation. We thank all study participants, whose feedback shaped EnRoute's development.


\small
\begin{thebibliography}{99}
\setlength{\itemsep}{1pt}

\bibitem{unep} United Nations Environment Programme. The most polluted capital in Europe you didn't even know about. \url{https://www.unep.org/news-and-stories/story/most-polluted-capital-europe-you-didnt-even-know-about}

\bibitem{dodovski} M. Dodovski and R. Temjanovski. 2020. Traffic in the City of Skopje and its impact on air pollution.

\bibitem{wri} S. Wang and M. Ge. 2019. Everything you need to know about the fastest-growing source of global emissions: Transport. World Resources Institute. \url{https://www.wri.org/blog/2019/10/everything-you-need-know-about-fastest-growing-source-global-emissions-transport}

\bibitem{planete} Planète Énergies. 2020. The global transportation sector: \COtwo{} emissions on the rise. \url{https://www.planete-energies.com/en/medias/close/global-transportation-sector-co2-emissions-rise}

\bibitem{niinimaki} K. Niinimäki, G. Peters, H. Dahlbo, P. Perry, T. Rissanen, and A. Gwilt. 2020. The environmental price of fast fashion. \emph{Nature Reviews Earth \& Environment} 1, 4, 189-200.

\bibitem{overconsumption} Overconsumption Project. Fast fashion: The economics and environmental cost of overconsumption. \url{https://overconsumption.org/blogs/news/fast-fashion-the-economics-and-environmental-cost-of-overconsumption/}

\bibitem{tsui} J. Tsui. 2020. The negative effects of corporate greenwashing. Sea Going Green. \url{https://www.seagoinggreen.org/blog/the-negative-effects-of-corporate-greenwashing}

\bibitem{disalvo} C. DiSalvo, P. Sengers, and H. Brynjarsdóttir. 2010. Mapping the landscape of sustainable HCI. In \emph{Proceedings of the SIGCHI Conference on Human Factors in Computing Systems (CHI '10)}. ACM, 1975-1984.

\bibitem{fogg} B. J. Fogg. 2003. \emph{Persuasive Technology: Using Computers to Change What We Think and Do}. Morgan Kaufmann.

\bibitem{consolvo} S. Consolvo, D. W. McDonald, and J. A. Landay. 2009. Theory-driven design strategies for technologies that support behavior change in everyday life. In \emph{Proceedings of CHI '09}. ACM, 405-414.

\bibitem{froehlich2010} J. Froehlich, L. Findlater, and J. Landay. 2010. The design of eco-feedback technology. In \emph{Proceedings of CHI '10}. ACM, 1999-2008.

\bibitem{ubigreen} J. Froehlich, T. Dillahunt, P. Klasnja, J. Mankoff, S. Consolvo, B. Harrison, and J. A. Landay. 2009. UbiGreen: Investigating a mobile tool for tracking and supporting green transportation habits. In \emph{Proceedings of CHI '09}. ACM, 1043-1052.

\bibitem{brynjarsdottir} H. Brynjarsdóttir, M. Håkansson, J. Pierce, E. Baumer, C. DiSalvo, and P. Sengers. 2012. Sustainably unpersuaded: How persuasion narrows our vision of sustainability. In \emph{Proceedings of CHI '12}. ACM, 947-956.

\bibitem{sandin} G. Sandin and G. M. Peters. 2018. Environmental impact of textile reuse and recycling: A review. \emph{Journal of Cleaner Production} 184, 353-365.

\bibitem{wrap} WRAP. 2012. \emph{Valuing Our Clothes: The True Cost of How We Design, Use and Dispose of Clothing in the UK}. Waste and Resources Action Programme.

\bibitem{iso14040} International Organization for Standardization. 2006. ISO 14040: Environmental management, life cycle assessment, principles and framework.

\bibitem{lin} X. Lin, X. Li, S. Kulkarni, and F. Zhao. 2021. The application of blockchain-based life cycle assessment on an industrial supply chain. \emph{Sustainability} 13, 23, 13332. \url{https://doi.org/10.3390/su132313332}

\bibitem{dw} Deutsche Welle. 2024. Targeting fast fashion: France aims to slow wasteful industry. \url{https://www.dw.com/en/targeting-fast-fashion-france-aims-to-slow-wasteful-industry/a-68731801}

\end{thebibliography}
\end{document}